%% file: classification.tex
\documentclass[a4paper,11pt]{article}

\usepackage{amsmath}
\usepackage{amssymb}
\usepackage[applemac]{inputenc}
\usepackage{latexsym}
\usepackage{graphicx}
\usepackage{hyperref}
\usepackage{color}
\usepackage[german, frenchb, english]{babel}

\newtheorem{thm}{Theorem}[section]
\newtheorem{prop}[thm]{Proposition}
\newtheorem{rem}[thm]{Remark}
\newtheorem{lem}[thm]{Lemma}
\newtheorem{cor}[thm]{Corollary}

\newtheorem{defe}[thm]{Definition}

\def\dem{{\hspace{-0.5 cm} $\Box$ \textbf{Proof of Proposition }}}
\def\demcor{{\hspace{-0.5 cm} $\Box$ \textbf{Proof of Corollary  }}}
\def\demthm{{\hspace{-0.5 cm} $\blacksquare$
\textbf{Proof of Theorem  }}}
\def\demlem{{\hspace{-0.5 cm} $\vartriangle$ \textbf{Proof of Lemma }}}

\def\cqfd{{\hfill $\Box$}}
\def\cqfdt{{\hfill $\blacksquare$}}
\def\cq{{\hfill $\vartriangle$}}

\def\C{{\mathbb{C}}}
\def\R{{\mathbb{R}}}

\def\N{{\mathbb{N}}}

\def\P{{\mathbb{P}}}
\def\K{{\mathbb{K}}}

\newcommand{\id}{{\rm id}}
\newcommand{\ad}{{\rm ad}}

\newcommand{\Gres}{{\rm Gr}_{\rm res}}
\newcommand{\Gr}{{\rm Gr}}

\newcommand{\Tr}{{\rm Tr}}

\newcommand{\orr}{{\rm or}}
\newcommand{\dimm}{{\rm dim}}
\newcommand{\U}{{\rm U}}
\newcommand{\Sp}{{\rm Sp}}

\newcommand{\Hc}{{\mathcal H}}
\newcommand{\Ze}{{\mathcal Z}}
\newcommand{\Le}{{\mathcal L}}

\newcommand{\g}{\mathfrak{g}}

\def\ha{{\mathfrak{h}}}
\def\k{{\mathfrak{k}}}

\begin{document}

\author{Alice Barbara TUMPACH\footnote{{\tt alice.tumpach@epfl.ch}, EPFL, Lausanne, Switzerland.
   This
  work was partially supported by
  the University of Paris VII, the University of
  Paris XI, and the \'Ecole Polytechnique, Palaiseau, France.}}

\date{}
\title{Classification of infinite-dimensional irreducible\\ Hermitian-symmetric  affine coadjoint
orbits} \maketitle

\abstract In the finite-dimensional setting, every
Hermitian-symmetric space of compact type is a coadjoint orbit of
a finite-dimensional Lie group. It is natural to ask whether every
infinite-dimensional Hermitian-symmetric space of compact type,
which is a particular example of an Hilbert manifold, is
transitively acted upon by a Hilbert Lie group of isometries.  In
this paper we give the classification of infinite-dimensional
irreducible Hermitian-symmetric  affine coadjoint orbits of
$L^{*}$-groups of compact type using the notion of simple roots of
non-compact type. The key step is, given an infinite-dimensional
symmetric pair $(\g, \k)$, where $\g$ is a simple $L^*$-algebra
and $\k$ a subalgebra of $\g$, to construct an increasing sequence
of finite-dimensional subalgebras $\g_{n}$ of $\g$ together with
an increasing sequence of finite-dimensional subalgebras $\k_{n}$
of $\k$ such that $\g = \overline{\cup\g_{n}}$, $\k =
\overline{\cup\k_{n}}$, and such that the pairs $(\g_{n}, \k_{n})$
are symmetric.
%Consequently the classification of infinite-dimensional
%irreducible Hermitian-symmetric  affine coadjoint orbits of
%$L^{*}$-groups can be deduced from the classification of
%finite-dimensional Hermitian-symmetric spaces.
Comparing with the classification of Hermitian-symmetric spaces
given by W.~Kaup, it follows that any Hermitian-symmetric space of
compact type is an affine-coadjoint orbit of an Hilbert Lie group.

%Five types occur~: the Grassmannians of finite-dimensional subspaces of a
%complex Hilbert space, the connected component containing
%$\Hc_{+}$ of the restricted Grassmannian of a polarized Hilbert
%space $\Hc = \Hc_{+}\oplus\Hc_{-}$, the Grassmannian of oriented
%$2$-planes in a real Hilbert space, the Grassmannian of
%orientation-preserving orthogonal complex structures close to a
%distinguished complex structure on a real Hilbert space, and the
%Grassmannian of Lagrangian subspaces of a symplectic Hilbert space
%close to a distinguished Lagrangian subspace.

%\tableofcontents
%\newpage

 %    \section{Classification of infinite-dimensional irreducible
  %   Hermitian-symmetric  affine coadjoint orbits}

\section{Introduction}

Let us introduce some notation. For any complex Hilbert space
$\mathcal{F}$ endowed with a distinguished basis $\{ f_{j}, j\in
J\}$, $\mathcal{F}_{\mathbb{R}}$ will denote the real Hilbert
space with basis $\{ f_{j}, j\in J\}$ and
$\mathcal{F}^{\mathbb{R}}$ the real Hilbert space with basis $\{
f_{j}, i f_j, j\in J\}$. The Hilbert space of Hilbert-Schmidt
operators on $\mathcal{F}$ will be denoted by $L^2(\mathcal{F})$,
and the Banach space of trace class operators on $\mathcal{F}$ by
$L^1(\mathcal{F})$.
 The group of invertible operators on $\mathcal{F}$ will be denoted by
$\textrm{GL}(\mathcal{F})$, and the group of unitary operators on
$\mathcal{F}$ by $\textrm{U}(\mathcal{F})$. In the sequel, $\Hc$
will denote a separable complex Hilbert space endowed with an
orthonormal basis $\{~e_{n}, n \in \mathbb{Z}\setminus\{0\}~\}$.
The Hermitian scalar product on $\Hc$ will be denoted by
$\langle\cdot\,,\,\cdot\rangle_{\Hc}$ and will be $\C$-skew-linear
with respect to the first variable, and $\C$-linear with respect
to the second variable. For a bounded operator $x$ on $\Hc$,
denote by $x^{T}$ the transpose of $x$ defined by $\langle x^{T}
e_{i}\,,\,e_{j}\rangle_{\Hc} = \langle x
e_{j}\,,\,e_{i}\rangle_{\Hc}$, and by $x^{*}$ the adjoint of $x$
defined by $\langle x^{*} e_{i}\,,\,e_{j}\rangle_{\Hc} = \langle
e_{i}\,,\,x e_{j}\rangle_{\Hc}$. The closed infinite-dimensional
subspace of $\Hc$ generated by the $e_{n}$'s for $n>0$ will be
called $\Hc_{+}$, and its orthogonal $\Hc_{-}$. For $0<p<+\infty$,
the $p$-dimensional subspace of $\Hc$ generated by the $e_{n}$'s
for $0<n\leq p$ will be denoted $\Hc_{p}$. Let $J_{0}$ be the
bounded operator on $\Hc$ defined by $J_{0} e_{i} = - e_{-i}$ if
$i< 0$ and $J_{0}e_{i} = e_{-i}$ if $i> 0$. For $\mathcal{F} =
\Hc, \Hc_{\pm}, \Hc_{p}$, or $\Hc_{p}^{\perp}$ define the
following Hilbert Lie groups and the associated Lie algebras
$$
\begin{array}{ll}
\textrm{GL}_{2}(\mathcal{F}) := \{g\in\textrm{GL}(\mathcal{F})\mid
g-\id\in
L^{2}(\mathcal{F})\}, & \mathfrak{gl}_{2}(\mathcal{F}) := L^{2}(\mathcal{F}),\\
\textrm{U}_{2}(\mathcal{F}) := \{g\in\textrm{U}(\mathcal{F})\mid
g-\id\in L^2(\mathcal{F})\},
& \mathfrak{u}_{2}(\mathcal{F}) := \{a\in L^{2}(\mathcal{F})\mid a^{*} + a = 0\},\\
\textrm{O}_{2}(\mathcal{F}_{\R}) :=
\{g\in\textrm{U}_{2}(\mathcal{F})\mid g^{T}g = \id\}, &
\mathfrak{o}_{2}(\mathcal{F}_{\R}) := \{a\in
L^{2}(\mathcal{F})\mid a^{T} + a = 0\}
\end{array}
$$
At last define
$$
\begin{array}{ll}
\textrm{Sp}_{2}(\Hc) := \{g\in\textrm{U}_{2}(\Hc)\mid g^{T}J_{0}g
= J_{0}\}, & \mathfrak{sp}_{2}(\Hc) := \{a\in L^{2}(\Hc)\mid
a^{T}J_{0} + J_{0}a = 0\}.
\end{array}
$$
On the Lie algebras $\g$ listed above, the bracket is the
commutator of operators and the Hermitian product
$\langle\cdot\,,\,\cdot\rangle_{\Hc}$ is defined using the trace
by
$$
\langle A, B \rangle := \Tr \,A^*B.
$$
These Lie algebras are $L^{*}$-algebras in the sense that the
following property is satisfied~:
$$ \langle [x\,,\,y]\,,\,z\rangle =
\langle y\,,\,[x^{*},\,z]\rangle $$ for every $x$, $y$ and $z$. In
fact, $\mathfrak{u}_{2}(\Hc)$,
$\mathfrak{o}_{2}(\Hc_{\mathbb{R}})$ and $\mathfrak{sp}_{2}(\Hc)$
are the unique infinite-dimensional simple $L^{*}$-algebras of
compact type modulo isomorphisms (see below for the corresponding
definition and \cite{Bal}, \cite{Har2}, or \cite{Uns2} for the
proof of this statement). The group associated to an
$L^{*}$-algebra is called an $L^{*}$-group. The $L^{*}$-groups
$\textrm{GL}_{2}(\Hc)$, $\textrm{U}_{2}(\Hc)$ and
$\textrm{Sp}_{2}(\Hc)$ are connected, but
$\textrm{O}_{2}(\Hc_{\mathbb{R}})$ admits two connected components
(see Proposition 12.4.2 on page 245 in \cite{PS}). The connected
component of $\textrm{O}_{2}(\Hc_{\R})$ containing the special
orthogonal group
$$
SO_{1}(\Hc_{\R}) := \{g\in\textrm{O}_{2}(\Hc_{\R})\mid g-\id\in
L^{1}(\Hc), \textrm{det}(g) = 1 \}
$$
will be denoted by $\textrm{O}^{+}_{2}(\Hc)$. The aim of this
paper is to prove the following statement.

\begin{thm}\label{orbiclas}
Every infinite-dimensional affine Hermitian-symmetric irreducible
coadjoint orbit of a simple $L^*$-group of compact type is
isomorphic to one of the following homogeneous space
\begin{enumerate}
\item the Grassmannian $\Gr^{(p)} = \U_{2}(\Hc) /
\left(\U_{2}(\Hc_{p}) \times U_{2}(\Hc_{p}^{\perp})\right)$ of
$p$-dimensional subspaces of $\Hc$ with $\dimm(\Hc_{p}) = p
<+\infty$

\item the connected component of the restricted Grassmannian
$\Gres^{0} = \U_{2}(\Hc) /
\left(\U_{2}(\Hc_{+})\times\U_{2}(\Hc_{-})\right)$ of the
polarized Hilbert space $\Hc = \Hc_{+} \oplus \Hc_{-}$ with
$\dimm\Hc_{+} = \dimm{\Hc}_{-} = +\infty$

\item the Grassmannian $\Gr^{(2)}_{\orr} = O^{+}_{2}(\Hc_{\R})/
\left(SO((\Hc_{2})_{\R}) \times
O^{+}_{2}((\Hc_{2})_{\R}^{\perp})\right)$ of oriented $2$-planes
in
  $\Hc_{\R}$,

\item the Grassmannian $\Ze(\Hc) = O^{+}_{2}(\Hc^{\R})/
U_{2}(\Hc)$ of orientation-preserving orthogonal complex
structures close to the distinguished complex structure on $\Hc$,

\item the Grassmannian  $\Le(\Hc) = \Sp_{2}(\Hc)/ \U_{2}(\Hc_{+})$
of Lagrangian subspaces close to $\Hc_{+}$.
\end{enumerate}
\end{thm}

In the finite-dimensional case, every Hermitian-symmetric space of
compact type is a coadjoint orbit of its connected group of
isometries (see Proposition~8.89 in \cite{Bes}). In the
infinite-dimensional setting, the biggest group of isometries of a
given Hermitian-symmetric space is not a Hilbert Lie group in
general. For example the restricted unitary group
$\U_{\text{res}}(\Hc)$ (see \cite{PS} for its definition) is a
Banach Lie group acting by isometries on the restricted
Grassmannian. It is a non trivial fact that the unitary Hilbert
Lie group $\U_{2}(\Hc)$, strictly contained in
$\U_{\text{res}}(\Hc)$, acts transitively on each connected
components of the restricted Grassmannian (see Proposition~5.2 in
\cite{BRT}). Theorem~\ref{orbiclas} above compared to the work of
W.~Kaup (\cite{Kau1}, \cite{Kau2}), leads to the following
generalization~:

\begin{cor}\label{coro}
Every Hermitian-symmetric space of compact type is an homogeneous
space of an Hilbert Lie group. More precisely, every
Hermitian-symmetric space of compact type is an affine-coadjoint
orbit of an $L^*$-group.
\end{cor}

   % \subsection{Orbites hermitiennes sym{\'e}triques}\label{r3}

%\~? Est-ce que les orbites affines sym{\'e}triques proviennent d'un
%$2$-cocycle continu.

%  En dimension finie, tout espace hermitien sym{\'e}trique compact est une
% orbite adjointe d'un groupe de Lie compact, le plus grand groupe
% d'isom{\'e}trie. Est-ce qu'a partir d'un espace hermitien-symetrique
% on peut reconstituer un groupe de Lie hilbertien qui agit
% transitivement sur l'espace~?~?~?

       %\subsection{G{\'e}n{\'e}ralit{\'e}s sur les espaces sym{\'e}triques}

 \section{Root Theory of complex $L^*$-algebra}\label{sysrac}

The root theory of complex $L^*$-algebras has been developed  by
J.\,R.~Schue in \cite{Schu1} and \cite{Schu2}.
%We will recall below
%some definitions and basic facts, and try to stress out the points
%that make the root theory of $L^*$-algebra \emph{so} similar to
%the root theory of finite-dimensional Lie algebras.
Let us first recall that an $L^*$-algebra  $\g$ over $\K\in\{\R,
\C\}$ is a Lie algebra over $\K$, which is also a Hilbert space
over $\K$ such that for every element  $x \in \g$, there exists
$x^{*} \in \g$ with the following property
\begin{equation}\label{*}
\langle [x, y], z \rangle = \langle y , [x^{*}, z] \rangle,
\end{equation}
for every  $y$, $z$ in $\g$. In the case when $\K = \C$, our
convention for the Hermitian product $\langle
\cdot\,,\cdot\rangle$ is that it is $\C$-skew-linear with respect
to the first variable, and $\C$-linear with respect to the second
variable. The first example of $L^*$-algebra is a semi-simple
finite-dimensional complex Lie algebra $\g_0$ endowed with an
involution $\sigma$, which defines a compact real form of $\g_0$.
In this example, the involutions $*$ and $\sigma$ are related by
$x^{*} = - \sigma(x)$ and the Hermitian scalar product is given by
$\langle x, y \rangle = B(x^{*}, y)$, where $B$ denotes the
Killing form of $\g_0$. An $L^*$-algebra is called of compact type
if $ x^{*} = -x $ for every $x$ in $\g$. For a given $L^*$-algebra
$\g$ the subspace
$$
\k:=\{x \in \g~|~x^* = -x\}
$$
is a real $L^*$-algebra of compact type. Thus an $L^*$-algebra can
be thought as an Hilbert Lie algebra together with a distinguished
compact real form.

For every subsets   $A$ and  $B$ of an $L^*$-algebra $\g$, $[A,
B]$ will denote the \emph{closure} of the vector space spanned by
 $\{ [a, b], a \in A, b \in B \}$. With this notation, an
 $L^*$-algebra is called semi-simple if  $\g = [\g,
\g]$, and simple if $\g$ is non-commutative and if every closed
ideal of $\g$ is trivial. Every $L^*$-algebra can be decomposed
into an orthogonal sum of its center and a semi-simple closed
ideal (see \cite{Tum}, 2.2.13.). A Cartan subalgebra of a complex
semi-simple $L^*$-algebra $\g^{\C}$ is defined as a maximal
Abelian $*$-stable subalgebra of $\g^{\C}$. Note that the
condition of being $*$-stable is added in comparison to the
finite-dimensional setting, hence a Cartan subalgebra may not be
maximal in the set of Abelian subalgebras. It is noteworthy that a
Cartan subalgebra of an $L^*$-algebra is in fact maximal Abelian
(see \cite{Schu2}, 1.1). Remark that a finite-dimensional Cartan
subalgebra of a complex semi-simple Lie algebra $\g^{\C}$ (for the
usual definition) is contained in a compact real form of
$\g^{\C}$, thus is also a Cartan subalgebra of the corresponding
finite-dimensional $L^*$-algebra.
 The existence of Cartan subalgebras of $L^*$-algebra is guarantied by
 Zorn's Lemma.
Every semi-simple $L^*$-algebra is an Hilbert sum of closed
$*$-stable simple ideals (see Theorem 1 in \cite{Schu1} for the
 initial proof. Some missing arguments can be found in \cite{Tum}).

In the sequel, $\g^{\C}$ will denote a semi-simple complex
$L^*$-algebra and $\ha^{\C}$ a Cartan subalgebra of $\g^{\C}$. A
root of $\g^{\C}$ with respect to $\ha^{\C}$ is defined, as in the
finite dimensional case, as an element $\alpha$ in the dual of
$\ha^{\C}$ such that the corresponding ``eigenspace''
$$
V_{\alpha}~:= \{ v \in \g^{\C} ~|~ \forall h \in \ha^{\C}, [h, v]
= \alpha(h) v  \}.
$$
is non-empty. In the following the set of non-zero roots with
respect to a given Cartan subalgebra will be denoted by
$\mathcal{R}$. Let us remark that a root has operator norm less
than $1$ and that for a non-zero root $\alpha$, the vector space
$V_{\alpha}$ is one-dimensional (see \cite{Schu1}). The Jacobi
identity implies that
 \begin{equation}\label{commutation}
[V_{\alpha}, V_{\beta}] \subset V_{\alpha + \beta}.
\end{equation}
By relation \eqref{*}, $V_{\alpha}^* = V_{-\alpha}$.
   The main achievement in \cite{Schu2} is to prove
  that a semi-simple complex $L^*$-algebra $\g^{\C}$ admits a Cartan
  Decomposition with respect to a given Cartan subalgebra $\ha^{\C}$ in
  the sense that $\g^{\C}$ is the Hilbert sum
\begin{equation}\label{cartandecomposition}
\g^{\C} = \ha^{\C} \oplus \sum_{\alpha \in \mathcal{R}}
V_{\alpha}.
\end{equation}
Let us remark that in a separable $L^*$-algebra, the set of root
is countable or finite.

By Zorn's Lemma, one can decompose the set $\mathcal{R}$ of
non-zero roots into  two disjoint subsets $\mathcal{R}_{+}$ and $
\mathcal{R}_{-}$ such that  $\alpha \in \mathcal{R}_{+}
\Leftrightarrow -\alpha \in \mathcal{R}_{-}$. Such a decomposition
defines a strict partial ordering on  $\mathcal{R}$ by
$$\alpha > \beta \Leftrightarrow \alpha - \beta > 0.$$
In the sequel, a decomposition $\mathcal{R} =
\mathcal{R}_{+}\cup\mathcal{R}_{-}$ as before and the induced
ordering  on the set of non-zero roots will be identified. The
elements in $\mathcal{R}_{+}$ will be called \emph{positive}
roots.

For every positive root  $\alpha$, one can choose $e_{\alpha} \in
V_{\alpha}$ such that  $\|e_{\alpha}\| = 1$. Then $e_{\alpha}^{*}
\in V_{-\alpha}$ and $\|e_{\alpha}^{*}\| = 1$. This choice made,
we define  $e_{\alpha}~:= e_{-\alpha}^{*}$ for $\alpha \in
\mathcal{R}_{-}$, in order to have, for every  $\alpha \in
\mathcal{R}$, the following relation $e_{\alpha}^{*} =
e_{-\alpha}$. By \eqref{cartandecomposition}, the set $\{
e_{\alpha}, \alpha \in \mathcal{R} \}$ is an Hilbert basis of
$\left(\ha^{\C}\right)^{\perp}$, and by \eqref{commutation},
$[e_{\alpha}, e_{\alpha}^{*}]$ belongs to $\ha^{\C}$. We define
the following elements in the Cartan subalgebra $\ha^{\C}$~:
\begin{equation}\label{ha}
h_{\alpha} := [e_{\alpha}, e_{\alpha}^{*}].
\end{equation}
A positive root is called \emph{simple} if it can not be written
as the sum of two positive roots. The set of simple roots will be
denoted by  $\mathcal{S}$. A subset $\mathcal{N}$ of the set of
non-zero roots $\mathcal{R}$ is called a \emph{root system}, if it
satisfies the following conditions~:
%\begin{equation}\label{rootsystem}
\begin{enumerate}
\item $\alpha \in \mathcal{N} \Rightarrow -\alpha \in
\mathcal{N}$, \item ($\alpha, \beta \in \mathcal{N}$  and  $\alpha
+ \beta \in \mathcal{R}) \Rightarrow \alpha + \beta \in
\mathcal{N}.$
\end{enumerate}
A subset $\mathcal{N}\subset\mathcal{R}$ is called
\emph{indecomposable} if it can not be written as the union of two
  orthogonal non-empty subsets. As in the classical theory, one
  has the following facts. The set $\mathcal{R}$ of non-zero roots
  of a simple $L^*$-algebra is indecomposable. If  $F$ is an indecomposable
  subset of the set of non-zero roots $\mathcal{R}$,
  then it generates a root system
$\mathcal{N}_{F}$, which is again indecomposable. The simple
$L^*$-algebra generated by $\{e_{\alpha}, \alpha \in
\mathcal{N}_{F}\}$ will be denoted by $\g(\mathcal{N}_{F})$.

For the classification of Hermitian-symmetric affine coadjoint
orbits given in next section, we will need the following results.
They were proved by J.\,R.~Schue in \cite{Schu1} in ordering to
classify the complex simple infinite-dimensional $L^*$-algebras.
%Propositions \ref{dec}, \ref{ordreinduit} and \ref{simple} will be
%en partie red\'emontr\'ees dans
%la preuve de la proposition \ref{bonordre} sur laquelle repose la
%classification des orbites coadjointes affines
%hermitiennes sym\'etriques irr\'eductibles
%des $L^{*}$-alg\`ebres simples de type compact.

\begin{prop}[\cite{Schu1}]\label{fini}
For every finite subset $F$  of the set of non-zero roots
$\mathcal{R}$ of a simple $L^*$-algebra, there exists a
\emph{finite} indecomposable system of non-zero roots containing
$F$.
\end{prop}

\begin{thm}[\cite{Schu1}, 3.2]\label{dec}
Let $\g^{\C}$ be a simple complex separable  $L^{*}$-algebra and
$\mathcal{R} = \{\alpha_{i}, i \in \N\setminus\{0\} \}$ the set of
non-zero roots with respect to a given Cartan subalgebra of
$\g^{\C}$. For every $n \in \N\setminus\{0\}$, set $F_{n}~:= \{
\alpha_{1}, \dots, \alpha_{n} \}$. Then there exists a sequence
$\{ \mathcal{N}_{n} \}_{n \in \N\setminus\{0\}}$ of \emph{finite}
subsets of $\mathcal{R}$ such that
\begin{enumerate}
\item $F_{n} \subset \mathcal{N}_{n} \subset \mathcal{N}_{n +
    1};$
\item $\mathcal{N}_{n}$ is a indecomposable root system; \item
$\mathcal{R} = \cup_{n \in \N\setminus\{0\}} \mathcal{N}_{n}$
\item the simple subalgebras $\g(\mathcal{N}_{n})$ generated by
$\mathcal{N}_{n}$ form a strictly increasing sequence with
$$
\g^{\C} = \overline{\cup_{n \in \N\setminus\{0\}}
\g(\mathcal{N}_{n})};
$$
\item The simple complex finite-dimensional algebras
$\g(\mathcal{N}_{n})$ are of the same Cartan type  A, B, C or D.
\end{enumerate}
\end{thm}

\begin{prop}[\cite{Schu1}, 3.2]\label{ordreinduit}
Given a sequence $\{\mathcal{N}_{n} \}_{n \in \N\setminus\{0\}}$
as in the previous Theorem, it is possible to define a
\emph{total} ordering on the vector space generated by the set of
roots such that~:
\begin{enumerate}
\item $\alpha > 0 \Rightarrow -\alpha < 0$; \item $\alpha > 0,
\beta > 0 \Rightarrow \alpha + \beta > 0$; \item If $\alpha > 0$
and $\alpha \notin \mathcal{N}_{n}$ then $\alpha
  > \beta$ for all $\beta \in \mathcal{N}_{n}$;
\item the induced ordering on $\mathcal{N}_{n}$ is a
lexicographical ordering  with respect to a basis of roots.
\end{enumerate}
\end{prop}

\begin{prop}[\cite{Schu1}, 3.3]\label{simple}
Let $\mathcal{S}$ be the set of simple roots of $\g^{\C}$ with
respect to the ordering defined in the previous Proposition. The
following assertions hold~:
\begin{enumerate}
\item $\mathcal{S}\,\cap\,\mathcal{N}_{n}$ is a complete system of
simple roots of the finite-dimensional algebra
$\g(\mathcal{N}_{n})$, i.e. every positive root
  $\alpha$
  of $\mathcal{N}_{n}$ can be written as a linear combination of elements in
  $\mathcal{S}\,\cap\,\mathcal{N}_{n}$ with non-negative integral coefficients; \item If
$\alpha$ and $\beta$ belong to $\mathcal{S}$, $\alpha
  - \beta$ is a root if and only if  $\alpha = \beta$;
\item the elements in $\mathcal{S}$ are linearly independent on
the reals
  and  every positive root $\alpha \in \mathcal{R}_{+}$ is a linear combination
  of elements in $\mathcal{S}$ with non-negative integral coefficients which are almost all
  zero.
%\item If $\tau = \sum_{i}n_{i} \alpha_{i}$ with $\alpha_{i} \in
%  \mathcal{S}$ and $n_{i} = 0$ sauf un nombre fini, alors pour
%savoir  si $\tau $ est une racine, il suffit de conna{\^\i}tre $\langle
 % h_{\alpha}, h_{\beta} \rangle$ pour tous $\alpha, \beta \in \mathcal{S}$.
\end{enumerate}
\end{prop}

% $k_{1}(\alpha, \beta) = k_{1}(\beta, \alpha) = 0$

       \section{Classification of irreducible Hermitian-symmetric affine coadjoint orbits}

Affine coadjoint orbits have been introduced in particular by
K.-H.~Neeb in \cite{Nee2}. The classification of
finite-dimensional Hermitian-symmetric coadjoint orbits using the
notion of roots of non-compact type has been carry out by
J.\,A.~Wolf in \cite{Wol2}. In the sequel, $\g$ will denote an
infinite-dimensional separable simple $L^{*}$-algebra of compact
type. According to \cite{Bal}, \cite{Har2} or \cite{Uns2}, it can
be realized as a subalgebra of the $L^*$-algebra
$\mathfrak{gl}_2(\Hc)$ consisting of Hilbert-Schmidt operators on
a separable complex Hilbert space $\Hc$. Let $G$ be the connected
$L^{*}$-group with Lie algebra $\g$, and $G^{\C}$ the connected
complex $L^*$-group with Lie algebra $\g^{\C} := \g \oplus i \g$.
By the duality $\g' = \g$ given by the trace, we can identify
affine adjoint and affine coadjoint orbits of $G$. Let
$\mathbb{D}$ be a derivation of $\g$ such that the affine
(co-)adjoint orbit $\mathcal{O}$ of
 $0$ in $\g$ associated to the affine adjoint action of $G$ defined
 by $\mathbb{D}$ is strongly K\"ahler (see \cite{Nee2}). By  Theorem~4.4 in  \cite{Nee2},
 there exists $D \in B(\Hc)$ satisfying $D^{*} = - D$ such that for all
 $x$ in
$\g$, $\mathbb{D}x = [D, x]$, as well as a Cartan subalgebra
$\mathfrak{h}^{\C}$ of  $\g^{\C}$  which is contained in $\ker
\mathbb{D}$. To emphasize the relation between the orbit and the
bounded operator $D$, we will often write $\mathcal{O} =
\mathcal{O}_{D}$. Abusing slightly the notation, we will sometimes
denote $\mathbb{D}$ by $\textrm{ad}(D)$. An alternative definition
of $\mathcal{O}_{D}$ is
$$
\mathcal{O}_{D} = \{ g D g^{-1} - D, g\in G\},
$$
and the affine adjoint action of $G$ on $\g$ is given by
$$
g\cdot a = \textrm{Ad}(g)(a) + g D g^{-1} - D
$$
where $g\in G$ and $a\in\g$. The subalgebra $\mathfrak{k}$ of $
\g$ which fixes $0$ is
$$
\mathfrak{k}~:= \{ x \in \g, [D, x] = 0 \}.
$$
It is an $L^*$-subalgebra of $\g$. Let $K$  be the associated
connected $L^*$-group.

The affine (co-)adjoint orbit $\mathcal{O}_{D}$ is called
\emph{(locally-)symmetric} if the orthogonal $\mathfrak{m}$ of
$\mathfrak{k}$ in $\g$ satisfies
$$
[\mathfrak{m}, \mathfrak{m}] \subset \k.
$$
The strongly K\"ahler orbit $\mathcal{O}_{D}$ is called
\emph{Hermitian-symmetric} if, for every $x$ in $\mathcal{O}_{D}$,
there exists a globally defined isometry ${s}_{x}$ (the
\emph{symmetry} with respect to $x$) preserving the complex
structure, such that $x$ is a fixed point of ${s}_{x}$, and such
that the differential of ${s}_{x}$ at $x$ is minus the identity of
$T_{x}\mathcal{O}_D$. An Hermitian-symmetric orbit is
(locally-)symmetric. In the following, we will be interested in
Hermitian-symmetric orbits $\mathcal{O}_{D}$. Let $\k^{\C}$ and
${\mathfrak{m}}^{\C}$ denote the complexifications of $\k$ and
$\mathfrak{m}$ respectively. Note that $\g^{\C}$ is the orthogonal
sum of $\k^{\C}$ and ${\mathfrak{m}}^{\C}$ with respect to the
Hermitian product of the $L^*$-algebra
 $\g^{\C}$.

\begin{prop}\label{kb}
Let $\mathfrak{h}^{\C}$ be a Cartan subalgebra of  $\g^{\C}$ that
is contained in   $\ker \ad D$, and let
$$
\g^{\C} = \mathfrak{h}^{\C} \oplus \sum_{\alpha \in \mathcal{R}}
V_{\alpha}
$$
be the associated Cartan decomposition of $\g^{\C}$, where
$\mathcal{R}$ denotes the set of non-zero roots with respect to
$\mathfrak{h}^{\C}$. Suppose that $\mathcal{O}_{D}$ is
Hermitian-symmetric. Then there exists two subsets $\mathcal{A}$
and $\mathcal{B}$ of $\mathcal{R}$ such that $\mathcal{A} \cup
\mathcal{B} = \mathcal{R}$ and
$$
\begin{array}{cc}
\k^{\C} = \mathfrak{h}^{\C} \oplus \sum_{\alpha \in \mathcal{A}}
V_{\alpha}, & \mathfrak{m}^{\C} = \sum_{\alpha \in \mathcal{B}}
V_{\alpha}.
\end{array}
$$
\end{prop}

\dem \ref{kb}~:\\
%%%%%%%%%%%%%%%%%%%%%
 Since $\mathcal{O}_{D}$ is (locally-)symmetric, one has
$\g^{\C} = \k^{\C} \oplus \mathfrak{m}^{\C}$ with
$$
[\k^{\C}, \k^{\C}] \subset \k^{\C}~;\qquad [\k^{\C},
\mathfrak{m}^{\C}] \subset \mathfrak{m}^{\C}~;\qquad
[\mathfrak{m}^{\C}, \mathfrak{m}^{\C}] \subset \k^{\C}.
$$
Let  $v$ be a non-zero vector in $V_{\alpha}$, and $v = v_{0} +
v_{1}$ his decomposition with respect to the direct sum  $\g^{\C}
= \k^{\C} \oplus \mathfrak{m}^{\C}$. For every $h \in \ha^{\C}$,
one has
$$
[h, v] = [h, v_{0} + v_{1}] = \alpha(h) (v_{0} + v_{1}) =
\alpha(h) v_{0} + \alpha(h) v_{1} = [h, v_{0}] + [h, v_{1}].
$$
Since $[\ha^{\C}, \k^{\C}] \subset \k^{\C}$ and $[ \ha^{\C},
  \mathfrak{m}^{\C}] \subset  \mathfrak{m}^{\C}$, it follows that
$$
\begin{array}{lll}
[h, v_{0}] = \alpha(h) v_{0} & \textrm{  et  } & [h, v_{1}] =
\alpha(h) v_{1}.
\end{array}
$$
But $V_{\alpha}$ is one-dimensional, hence either  $v_{0} = 0$, or
$v_{1} = 0$. Consequently $V_{\alpha}$ is contained either in
$\k^{\C}$ or in $\mathfrak{m}^{\C}$. \cqfd

\begin{prop}\label{DV}
For every $\alpha \in \mathcal{R}$, there exists a constant
$c_{\alpha} \in \R$ such that $[D, e_{\alpha}] = i c_{\alpha}\,
e_{\alpha}$. Moreover $c_{-\alpha} = - c_{\alpha}$.
\end{prop}

\dem \ref{DV}~:\\
For every  $\alpha \in \mathcal{R}$ and every  $h \in \ha^{\C}$,
one has
$$
[h, [D, e_{\alpha}]] = [[h, D], e_{\alpha}] +[D, [h, e_{\alpha}]]
= \alpha(h) \,[D, e_{\alpha}].
$$
The space $V_{\alpha}$ being one-dimensional, it follows that $[D,
e_{\alpha}]$ is proportional to $e_{\alpha}$. Since $D$ satisfies
$D^{*} = - D$, one has, for every $\alpha \in \mathcal{R}$, the
following relation
$$
\langle [D, e_{\alpha}], e_{\alpha} \rangle = - \langle e_{\alpha}
, [D, e_{\alpha}] \rangle = - \overline{\langle [D,
  e_{\alpha}], e_{\alpha} \rangle }.
$$
Thus there exists a real constant $c_{\alpha}$ such that
$$ [D,
e_{\alpha}] = i c_{\alpha} \,e_{\alpha}
$$
On the other hand,
$$
[ D, e_{\alpha}]^{*} = [e_{\alpha}^{*}, D^{*}] =
 - [e_{\alpha}^{*}, D] = [D, e_{\alpha}^{*}].
$$
Whence
$$
\langle [D, e_{\alpha}^{*}], e_{\alpha}^{*} \rangle = \langle
e_{\alpha}, [D, e_{\alpha}^{*}]^{*} \rangle =  \langle e_{\alpha},
[ D, e_{\alpha}] \rangle = i c_{\alpha}.
$$
Consequently $[D, e_{\alpha}^{*}] = - i c_{\alpha} e_{\alpha}^*$.
\cqfd

\begin{rem} {\rm
Let us denote by  $\mathfrak{m}_{+}$ (resp. $\mathfrak{m}_{-}$)
the closed subspace of $\g^{\C}$ generated by the $e_{\alpha}$'s,
where
  $\alpha$ runs over the set of roots for which  $c_{\alpha} > 0$
  (resp. $c_{\alpha} < 0$). Let $\mathcal{B}_{+}$
  (resp. $\mathcal{B}_{-}$) be the set of roots $\beta$ in $\mathcal{B}$
  such that $V_{\beta} \in
  \mathfrak{m}_{+}$ (resp. $V_{\beta} \in \mathfrak{m}_{-}$).
 }
\end{rem}

\begin{defe} {\rm
The affine adjoint orbit  $\mathcal{O}_{D}$ is called
\emph{(isotropy-)irreducible} if $\mathfrak{m}$ is a non-zero
irreducible $\textrm{Ad}(K)$-module. }
\end{defe}

\begin{prop}\label{b}
If the affine adjoint orbit  $\mathcal{O}_{D}$ is irreducible,
then $\mathfrak{m}_{+}$ and $\mathfrak{m}_{-}$ are irreducible
 $\textrm{Ad(K)}$-modules, and there exists a constant
$c > 0$ such that $\ad(D)_{|\mathfrak{m}_{+}} = i c\,\,
\id_{|\mathfrak{m}_{+}}$ and $\ad(D)_{|\mathfrak{m}_{-}} = -i
c\,\, \id_{|\mathfrak{m}_{-}}$. In particular, the spectrum of
$\ad(D)$ is $\{ 0, ic, -ic \}$, hence $D$ admits exactly two
distinct eigenvalues.
\end{prop}

\dem \ref{b}~:\\
For every $k \in \k$ and every $e_{\alpha} \in
\mathfrak{m}_{\pm}$, one has
$$
[D, [k, e_{\alpha}]] = [[D, k], e_{\alpha}] + [k, [D, e_{\alpha}]]
= i c_{\alpha} [k, e_{\alpha}].
$$
It follows that $[\k, \mathfrak{m}_{\pm}] \subset
\mathfrak{m}_{\pm}$ and that $\mathfrak{m}_{\pm}$ is stable under
the adjoint action of   $K$. Let us suppose that
$\mathfrak{m}_{+}$ decomposes into a sum of two non-zero
$\textrm{Ad(K)}$-modules $\mathfrak{m}_{1}$ and
$\mathfrak{m}_{2}$. Then
$$
\mathfrak{m}_{-} = \mathfrak{m}_{1}^{*} \oplus
\mathfrak{m}_{2}^{*},
$$
and it follows that $\mathfrak{m}$ decomposes also into the sum of
two non-zero $\textrm{Ad(K)}$-modules, namely $\g \cap
(\mathfrak{m}_{1} \oplus \mathfrak{m}_{1}^{*})$ and $\g \cap
(\mathfrak{m}_{2} \oplus \mathfrak{m}_{2}^{*})$. The orbit
$\mathcal{O}_{D}$ being irreducible, $\mathfrak{m}$ is an
irreducible  $\textrm{Ad}(K)$-module and this leads to a
contradiction. So the irreducibility of $\mathfrak{m}_{\pm}$ is
proved. Let $e_{\alpha}$ be an element in $\mathfrak{m}_{+}$ and
set $c = c_{\alpha}$~:
$$
[ D, e_{\alpha}] = i c \,e_{\alpha}.
$$
The kernel $\ker(D - i c)$ being an $\textrm{Ad}(K)$-module of
$\mathfrak{m}_{+}$, one has $\textrm{ad}(D)_{|\mathfrak{m}_{+}} =
i c \textrm{ id}_{|\mathfrak{m}_{+}}$. The relation $c_{- \alpha}
= - c_{\alpha}$ implies that $\textrm{ad}(D)_{|\mathfrak{m}_{-}} =
- i c \textrm{ id}_{|\mathfrak{m}_{-}}$. \cqfd

\begin{defe} {\rm
Given an ordering on the set of non-zero roots $\mathcal{R}$ of
$\g^{\C}$, a simple roots $\phi$ is called of \emph{non-compact
type} (see \cite{Wol2}) if every root $\alpha \in \mathcal{R}$ is
of the form
$$
\alpha = \pm \sum_{\Psi \in \mathcal{S}-\{\phi\}} a_{\Psi} \Psi,
a_{\Psi} \geq 0,
$$
or of the form
$$
\alpha = \pm \left(\phi +  \sum_{\Psi \in \mathcal{S}-\{\phi\}}
a_{\Psi} \Psi\right),   a_{\Psi} \geq 0.
$$
} \end{defe}

\begin{lem} \label{ordrebeta}
Let $\mathcal{O}_{D}$ be a Hermitian-symmetric affine adjoint
irreducible orbit of a simple $L^{*}$-algebra $\g$,  $\ha^{\C}$ be
a Cartan subalgebra of  $\g^{\C}$ contained in  $\ker \ad D$, and
$$
\g^{\C} = \mathfrak{h}^{\C} \oplus \sum_{\alpha \in \mathcal{A}}
V_{\alpha} \oplus \sum_{\beta \in \mathcal{B}} V_{\beta}
$$
be the associated Cartan decomposition of $\g^{\C}$ with
$$
\mathfrak{k}^{\C} = \mathfrak{h}^{\C} \oplus \sum_{\alpha \in
\mathcal{A}} V_{\alpha}, \textrm{    and    } \mathfrak{m}^{\C} =
\sum_{\beta \in \mathcal{B}} V_{\beta}.
$$
For every ordering  $\mathcal{R} = \mathcal{R}_{+} \cup
\mathcal{R}_{-}$ on the set of roots, there exists a unique simple
root $\phi$ belonging to $\mathcal{B}$.
\end{lem}

\demlem \ref{ordrebeta}:

 Let $\{\phi_{i}, \Psi_{j}\}_{i \in I, j \in J}$ be the set of
 simple roots with
 $\phi_{i}$  in $\mathcal{B}$ and $\Psi_{j}$ in
$\mathcal{A}$. Let us suppose that $I$ is empty. The relation
$[\k^{\C}, \k^{\C}] \subset \k^{\C}$ implies that every positive
root belongs to  $\mathcal{A}$ and consequently $\mathfrak{m} =
\{0\}$, which contradicts the hypothesis that $\mathfrak{m}$ is a
non-zero irreducible  $\textrm{Ad}(K)$-module. Let $\phi$ be a
simple root in $\mathcal{B}$. The closed vector space spanned by
the adjoint action of  $\mathfrak{k}$ on $e_{\phi}$ is a non-zero
irreducible  $\textrm{Ad}(K)$-module of $\mathfrak{m}^{\C}$. It
follows that $\phi$ is necessarily unique. \cq

\begin{lem} \label{noncom}
Under the hypothesis of Lemma~\ref{ordrebeta}, there exists an
increasing sequence of finite indecomposable root systems
$\mathcal{N}_{n}$ such that
\begin{enumerate}
\item \label{1} $\mathcal{R} = \cup_{n \in \N\setminus\{0\}}
\mathcal{N}_{n}$; \item \label{2} all the finite-dimensional
subalgebras  $\g(\mathcal{N}_{n})$ generated by
  $\mathcal{N}_{n}$ belong to the same type A, B, C, or D
  and $\g^{\C}$ is the closure of the union of the subalgebras $\g(\mathcal{N}_{n})$;
\item \label{3} $\phi$ is a simple root of non-compact type for
each subalgebra
  $\g(\mathcal{N}_{n})$ with respect to the  ordering on the roots of $\g(\mathcal{N}_{n})$
  induced by the ordering on  $\mathcal{R}$ defined in Proposition \ref{ordreinduit}.
%\item il existe une suite croissante $\{ k_{n}, n \in \N\setminus\{0\}\}$
%de sous-alg\`ebres de Lie de  $\mathfrak{k}$
\end{enumerate}
\end{lem}

\demlem \ref{noncom}:\\
Let $\{\alpha_{1}, \dots, \alpha_{n}, \dots\}$ be a numbering of
the roots in $\mathcal{A}$. Set $F_{n} = \{\alpha_{1}, \dots,
\alpha_{n}\}$. Let us construct by induction an increasing
sequence of finite indecomposable root systems  $\mathcal{N}_{n}$
as follows. By Proposition \ref{fini}, there exists a finite
indecomposable root system $\mathcal{N}_{1}$ containing $\{\phi\}
\cup F_{1}$. Suppose that $\mathcal{N}_{n-1}$ is constructed, then
there exists a finite indecomposable root system $\mathcal{N}_{n}$
containing $F_{n} \cup \mathcal{N}_{n-1}$. Since every root in
$\mathcal{B}$ is the sum of $\phi$ and roots in $\mathcal{A}$,
$\mathcal{R} = \cup_{n \in \N\setminus\{0\}} \mathcal{N}_{n}$. The
sequence of finite-dimensional simple subalgebras
$\g(\mathcal{N}_{n})$ generated by the root systems
$\mathcal{N}_{n}$ is increasing and such that  $\g^{\C} =
\overline{\cup_{n
  \in \N\setminus\{0\}} \g(\mathcal{N}_{n})}$.
 Since there exists only  9 types of finite-dimensional simple
 algebras, at least one type occurs an infinite number of times. Since $\g^{\C}$
 is  infinite-dimensional and since only the types
 A, B, C, or D
corresponds to algebras of arbitrary dimension, at least one of
the types A, B, C, or D occurs an infinite number of times. It
follows that there exists a subsequence $\mathcal{N}_{n_{k}}$ of
$\mathcal{N}_{n}$
 such that all the subalgebras
$\g(\mathcal{N}_{n_{k}})$ are of the same type A, B, C, or D. Let
$\mathcal{S}_{n_{k}}$ be the set of simple roots of
$\g(\mathcal{N}_{n_{k}})$ with respect to the ordering induced by
the ordering on $\mathcal{R}$ defined in
Proposition~\ref{ordreinduit}. By Proposition~\ref{simple},
$\mathcal{S}_{n_{k}} = \mathcal{S} \cap \g(\mathcal{N}_{n_{k}})$,
where $\mathcal{S}$ is the set of simple roots of $\g^{\C}$. For
every positive root $\gamma$ in $\mathcal{N}_{n_{k}}$, there
exists a finite sequence $\{\gamma_{i}, i = 1, \dots, k\}$ of
roots in $\mathcal{S}_{n_{k}}$ such that
$$
\gamma = \gamma_{1} + \gamma_{2} + \dots + \gamma_{k},
$$
and such that the partial sums  $\gamma_{1} + \dots + \gamma_{j}$,
$1 \leq j \leq k$ are roots (see \cite{Bou}). Hence the vector
space
 $V_{\gamma}$ is generated by
$$
v = [ e_{\gamma_{k}}, [e_{\gamma_{k-1}},
    [e_{\gamma_{k-2}}, \dots, [e_{\gamma_{2}},
    e_{\gamma_{1}}] \dots
]]].
$$
The orbit $\mathcal{O}_{D}$ being irreducible,  $[D, e_{\phi}] =
\epsilon_{\phi}\, i c \,e_{\phi} $ with  $\epsilon_{\phi} = +1$
(resp. $-1$) if $V_{\phi} \subset \mathfrak{m}_{+}$ (resp.
$\mathfrak{m}_{-}$). Whence
$$
[D, v ] = \textrm{card}\left(\{i, \gamma_{i} = \phi\}\right)
\epsilon_{\phi}\, i c \,v.
$$
Since $\textrm{ad}(D)$ preserves $\mathfrak{k}^{\C}$,
$\mathfrak{m}_{+}$ and $\mathfrak{m}_{-}$, it follows that for
$\gamma$  in $\mathcal{A} \cap \mathcal{R}_{+}$,
$\textrm{card}\left(\{i, \gamma_{i} = \phi\}\right) = 0$ and for
$\gamma$ in $\mathcal{B} \cap \mathcal{R}_{+}$,
$\textrm{card}\left(\{i, \gamma_{i} = \phi \}\right) = 1$.
Consequently $\phi$ is of non-compact type. \cq

\begin{prop}\label{suiteespsym}
Let $\mathcal{O} = G/K$ be a Hermitian-symmetric irreducible
affine coadjoint orbit of an $L^{*}$-group $G$ of compact type,
and $\g = \mathfrak{k} \oplus \mathfrak{m}$ the associated
decomposition of the Lie algebra $\g$ of $G$, where $\mathfrak{k}$
is the Lie algebra of the isotropy group $K$. Then there exists an
increasing sequence of finite-dimensional subalgebras $\g_{n}$ of
$\g$, of the same type
 A, B, C or D, and an increasing sequence of subalgebras
$\mathfrak{k}_{n}$ of $\mathfrak{k}$
 such that
\begin{enumerate}
\item $\g = \overline{\cup \g_{n}}$ \item $\mathfrak{k} =
\overline{\cup \k_{n}}$ \item for every $n \in \N\setminus\{0\}$,
the orthogonal $\mathfrak{m}_{n}$ of $\mathfrak{k}_{n}$ in
  $\g_{n}$ satisfies
  $$
[\mathfrak{k}_{n}, \mathfrak{m}_{n}] \subset
\mathfrak{m}_{n}\qquad\textrm{and}\qquad [\mathfrak{m}_{n},
\mathfrak{m}_{n}] \subset  \mathfrak{k}_{n},
$$
hence $(\g_{n}, \mathfrak{k}_{n})$ is a symmetric pair.
\end{enumerate}
\end{prop}

\dem \ref{suiteespsym}~:\\
This is a direct consequence of Lemma \ref{noncom}, with $\g_{n} =
\g \cap \g(\mathcal{N}_{n})$ and $\mathfrak{k}_{n} = \mathfrak{k}
\cap \g(\mathcal{N}_{n})$. \cqfd
\\

From the discussion above it follows that the classification of
Hermitian-symmetric irreducible  affine coadjoint orbits of
$L^*$-groups of compact type can be deduced from the knowledge of
the simple roots of non-compact type of finite-dimensional simple
algebras (see the proof of Theorem~\ref{orbiclas} below). A simple
root of a simple finite-dimensional algebra is of non-compact type
if and only if it appears with the coefficient $+1$ in the
expression of the greatest root. We recall the list of simple
roots of non-compact type in the finite-dimensional Lie algebras
of type A, B, C, or D in tabular \ref{rnc} (see \cite{Lie} or
\cite{Wol2}).
\\

\newpage

%\begin{figure} \centering \input{algebres.pstex_t} \caption{Voici une
%grassmannienne de dimension infinie. \label{grass}} \end{figure}
\begin{table}[h, width = 2cm] \centering
\input{rnce.pstex_t}
\caption{{\it Simple roots of non-compact type in the simple
finite-dimensional Lie algebras of type
   A, B, C and D.\label{rnc}}}
\end{table}

\demthm \ref{orbiclas}~:\\
By Lemma~\ref{ordrebeta}, there exists a unique simple root $\phi$
in $\mathcal{B}$ regardless to the ordering chosen on the set of
non-zero roots $\mathcal{R}$. By Lemma~\ref{noncom} part 3.,
$\phi$ is a simple root of non-compact type for each
finite-dimensional subalgebras $\g(\mathcal{N}_{n})$ constructed
in Lemma~\ref{noncom} part 2. when $\mathcal{R}$ is endowed with
the particular ordering constructed in
Proposition~\ref{ordreinduit}. For this ordering, simple roots of
$\g(\mathcal{N}_{n})$ are simple roots of $\g^{\C}$. It follows
that the set of possible roots $\phi$ can be deduced from tabular
\ref{rnc}. Such a root $\phi$ defines a symmetric pair $(\g, \k)$
where $\k$ is the $L^{*}$-algebra of compact type whose Dynkin
diagram is obtained by removing $\phi$ from the Dynkin diagram of
$\g$ ($\k$ is the orthogonal of the vector space generated by the
$e_{\phi + \alpha}$'s). \cqfdt
\\
\\
\demcor \ref{coro}~:\\
In \cite{Kau2}, W.~Kaup gives the classification of
Hermitian-symmetric spaces of arbitrary dimension using the theory
of Hermitian Jordan-triple system developed in the Banach context
in \cite{Kau1}. It is straightforward to verify that both lists
coincide.\cqfd

\begin{rem}{\rm
The Grassmannian $\Gr^{(p)} = \U_{2}(\Hc) / \left(\U_{2}(\Hc_{p})
\times U_{2}(\Hc_{p}^{\perp})\right)$ of $p$-dimensional subspaces
of $\Hc$ with $\dimm(\Hc_{p}) = p <+\infty$, is the affine adjoint
orbit of $\U_{2}(\Hc)$ for the derivations defined by the bounded
operators $D_{k, l}^{(p)} = i k\, \textrm{p}_{\Hc_{p}} - i l\,
\textrm{p}_{\Hc_{p}^{\perp}}$, where $k, l \in \mathbb{R}$, $k\neq
-l$, and $\textrm{p}_{\Hc_{p}}$ (resp.
$\textrm{p}_{\Hc_{p}^{\perp}}$) is the orthogonal projection onto
$\Hc_{p}$ (resp. $\Hc_{p}^{\perp}$). The homogeneous space
$\Gr^{(p)}$ is therefore endowed with a one-parameter family of
K\"ahler structure (encoded by $(k+l)$). The derivation $D_{k,
l}^{(p)}$ is inner if and only if $l = 0$. For $p = 1$,
$\Gr^{(p)}$ is the projective space of $\Hc$.

The restricted Grassmannian $\Gres$ has been studied in \cite{PS}
and \cite{Wur1}. The connected component $\Gres^{0}$ of $\Gres$
containing $\Hc_+$ is the affine adjoint orbit of $\U_{2}(\Hc)$
for the derivations defined by the bounded operators $D_{k,
l}^{(\infty)} = i k \,\textrm{p}_{+} - i l \,\textrm{p}_{-}$,
where $k, l \in \mathbb{R}$, $k\neq -l$, and $\textrm{p}_{\pm}$ is
the orthogonal projection onto $\Hc_{\pm}$. None of these
derivations is inner.

The Grassmannian $\Gr^{(2)}_{\orr} = O^{+}_{2}(\Hc_{\R})/
\left(SO((\Hc_{2})_{\R}) \times
O^{+}_{2}((\Hc_{2})_{\R}^{\perp})\right)$ of oriented $2$-planes
in $\Hc_{\R}$ is the $O^{+}_{2}(\Hc_{\R})$-adjoint orbit of $kJ$
where $k \neq 0$ and $J$ is the natural complex structure on
$(\Hc_{2})_{\mathbb{R}}$. It can be identified as complex manifold
with the conic $\mathcal{C}$ in the complex projective space
$\mathbb{P}(\Hc)$ defined by
$$
\mathcal{C} := \{(z_{1} : \dots : z_{n}: \dots) \in \P(\Hc),
\sum_{i
  \in \N} z_{i}^{2} = 0 \}.
$$

Denote by $(\cdot, \cdot)$ the real part of the Hermitian scalar
product on $\Hc$. The Grassmannian $\Ze(\Hc) =
O^{+}_{2}(\Hc^{\R})/ U_{2}(\Hc)$ is the space of complex
structures $I$ on $\Hc^{\R}$ such that
$$
(IX, IY) = (X, Y),
$$
defining the same orientation as the distinguished complex
structure $I_0$ on $\Hc$ and being closed to it. For every $k \neq
0$, the space $\Ze(\Hc)$
  can be identified with the $O^{+}_{2}(\Hc^{\R})$-affine adjoint orbit
  of 0
for the the bounded operator $D_k^{(0)} = kI_{0}$. Denote by
$\Hc^\C$ the $\C$-extension of $\Hc^{\R}$ and by $Z_{+}$ (resp.
$Z_{-}$) the eigenspace of the $\C$-linear extension of $I_0$ with
eigenvalue $+i$ (resp. $-i$). One has $\Hc^{\C} = Z_+ \oplus
Z_-$. The homogeneous space $\Ze(\Hc)$ injects into the restricted
Grassmannian  of the polarized Hilbert space $\Hc^{\C} = Z_{+}
\oplus  Z_{-}$ as a totally geodesic submanifold via the
application which maps a complex structure $I$ to the subspace
 consisting of $(1, 0)$-type vectors $X$ with
respect to $I$, i.e. satisfying $IX = iX.$

The Grassmannian  $\Le(\Hc) = \Sp_{2}(\Hc)/ \U_{2}(\Hc_{+})$ of
Lagrangian subspaces close to $\Hc_{+}$ is the
$\Sp_{2}(\Hc)$-affine
  adjoint orbit of $0$ for the derivations given by the bounded operators $D_{l,l}^{(\infty)} = i l\,\textrm{p}_{+} -
  i\l
  \textrm{p}_{-}$. It is a totally geodesic
  submanifold
  of the restricted Grassmannian
  $\Gres$.
}
\end{rem}

\section*{Acknowledgments}
Our thanks go to J.A.~Wolf for very pleasant discussions.

\end{document}

%% file: rnce.pstex_t
\begin{picture}(0,0)%
\includegraphics{rnce.pstex}%
\end{picture}%
\setlength{\unitlength}{3947sp}%
\begingroup\makeatletter\ifx\SetFigFont\undefined%
\gdef\SetFigFont#1#2#3#4#5{%
  \reset@font\fontsize{#1}{#2pt}%
  \fontfamily{#3}\fontseries{#4}\fontshape{#5}%
  \selectfont}%
\fi\endgroup%
\begin{picture}(3010,3684)(-283,-4670)
\put(-221,-2143){\makebox(0,0)[lb]{\smash{{\SetFigFont{7}{8.4}{\rmdefault}{\mddefault}{\updefault}{\color[rgb]{0,0,0}Type B:}%
}}}}
\put(969,-4292){\makebox(0,0)[lb]{\smash{{\SetFigFont{7}{8.4}{\rmdefault}{\mddefault}{\updefault}{\color[rgb]{0,0,0}$\alpha_{4}$}%
}}}}
\put(1794,-4292){\makebox(0,0)[lb]{\smash{{\SetFigFont{7}{8.4}{\rmdefault}{\mddefault}{\updefault}{\color[rgb]{0,0,0}$\alpha_{n-2}$}%
}}}}
\put(145,-4434){\makebox(0,0)[lb]{\smash{{\SetFigFont{7}{8.4}{\rmdefault}{\mddefault}{\updefault}{\color[rgb]{0,0,0}$\alpha_{2}$}%
}}}}
\put(-268,-4179){\makebox(0,0)[lb]{\smash{{\SetFigFont{6}{7.2}{\rmdefault}{\bfdefault}{\updefault}{\color[rgb]{0,0,0}Type D:}%
}}}}
\put(-221,-1196){\makebox(0,0)[lb]{\smash{{\SetFigFont{6}{7.2}{\rmdefault}{\bfdefault}{\updefault}{\color[rgb]{0,0,0}Type A:}%
}}}}
\put(-221,-3067){\makebox(0,0)[lb]{\smash{{\SetFigFont{6}{7.2}{\rmdefault}{\bfdefault}{\updefault}{\color[rgb]{0,0,0}Type C:}%
}}}}
\put(2115,-4056){\makebox(0,0)[lb]{\smash{{\SetFigFont{7}{8.4}{\rmdefault}{\mddefault}{\updefault}{\color[rgb]{0,0,0}$\alpha_{n-1}$}%
}}}}
\put(1382,-4056){\makebox(0,0)[lb]{\smash{{\SetFigFont{7}{8.4}{\rmdefault}{\mddefault}{\updefault}{\color[rgb]{0,0,0}$\alpha_{5}$}%
}}}}
\put(145,-3867){\makebox(0,0)[lb]{\smash{{\SetFigFont{7}{8.4}{\rmdefault}{\mddefault}{\updefault}{\color[rgb]{0,0,0}$\alpha_{1}$}%
}}}}
\put(603,-4056){\makebox(0,0)[lb]{\smash{{\SetFigFont{7}{8.4}{\rmdefault}{\mddefault}{\updefault}{\color[rgb]{0,0,0}$\alpha_{3}$}%
}}}}
\put(237,-3255){\makebox(0,0)[lb]{\smash{{\SetFigFont{7}{8.4}{\rmdefault}{\mddefault}{\updefault}{\color[rgb]{0,0,0}$\alpha_{1}$}%
}}}}
\put(924,-3255){\makebox(0,0)[lb]{\smash{{\SetFigFont{7}{8.4}{\rmdefault}{\mddefault}{\updefault}{\color[rgb]{0,0,0}$\alpha_{3}$}%
}}}}
\put(1657,-3255){\makebox(0,0)[lb]{\smash{{\SetFigFont{7}{8.4}{\rmdefault}{\mddefault}{\updefault}{\color[rgb]{0,0,0}$\alpha_{n-2}$}%
}}}}
\put(2527,-3255){\makebox(0,0)[lb]{\smash{{\SetFigFont{7}{8.4}{\rmdefault}{\mddefault}{\updefault}{\color[rgb]{0,0,0}$\alpha_{n}$}%
}}}}
\put(2024,-2923){\makebox(0,0)[lb]{\smash{{\SetFigFont{7}{8.4}{\rmdefault}{\mddefault}{\updefault}{\color[rgb]{0,0,0}$\alpha_{n-1}$}%
}}}}
\put(1244,-2923){\makebox(0,0)[lb]{\smash{{\SetFigFont{7}{8.4}{\rmdefault}{\mddefault}{\updefault}{\color[rgb]{0,0,0}$\alpha_{n-3}$}%
}}}}
\put(1657,-2312){\makebox(0,0)[lb]{\smash{{\SetFigFont{7}{8.4}{\rmdefault}{\mddefault}{\updefault}{\color[rgb]{0,0,0}$\alpha_{n-2}$}%
}}}}
\put(237,-2312){\makebox(0,0)[lb]{\smash{{\SetFigFont{7}{8.4}{\rmdefault}{\mddefault}{\updefault}{\color[rgb]{0,0,0}$\alpha_{1}$}%
}}}}
\put(924,-2312){\makebox(0,0)[lb]{\smash{{\SetFigFont{7}{8.4}{\rmdefault}{\mddefault}{\updefault}{\color[rgb]{0,0,0}$\alpha_{3}$}%
}}}}
\put(2527,-2312){\makebox(0,0)[lb]{\smash{{\SetFigFont{7}{8.4}{\rmdefault}{\mddefault}{\updefault}{\color[rgb]{0,0,0}$\alpha_{n}$}%
}}}}
\put(2068,-2030){\makebox(0,0)[lb]{\smash{{\SetFigFont{7}{8.4}{\rmdefault}{\mddefault}{\updefault}{\color[rgb]{0,0,0}$\alpha_{n-1}$}%
}}}}
\put(1291,-2030){\makebox(0,0)[lb]{\smash{{\SetFigFont{7}{8.4}{\rmdefault}{\mddefault}{\updefault}{\color[rgb]{0,0,0}$\alpha_{n-3}$}%
}}}}
\put(512,-2030){\makebox(0,0)[lb]{\smash{{\SetFigFont{7}{8.4}{\rmdefault}{\mddefault}{\updefault}{\color[rgb]{0,0,0}$\alpha_{2}$}%
}}}}
\put(557,-2923){\makebox(0,0)[lb]{\smash{{\SetFigFont{7}{8.4}{\rmdefault}{\mddefault}{\updefault}{\color[rgb]{0,0,0}$\alpha_{2}$}%
}}}}
\put(1017,-1368){\makebox(0,0)[lb]{\smash{{\SetFigFont{7}{8.4}{\rmdefault}{\mddefault}{\updefault}{\color[rgb]{0,0,0}$\alpha_{3}$}%
}}}}
\put(1474,-1085){\makebox(0,0)[lb]{\smash{{\SetFigFont{7}{8.4}{\rmdefault}{\mddefault}{\updefault}{\color[rgb]{0,0,0}$\alpha_{n-3}$}%
}}}}
\put(1794,-1368){\makebox(0,0)[lb]{\smash{{\SetFigFont{7}{8.4}{\rmdefault}{\mddefault}{\updefault}{\color[rgb]{0,0,0}$\alpha_{n-2}$}%
}}}}
\put(2206,-1085){\makebox(0,0)[lb]{\smash{{\SetFigFont{7}{8.4}{\rmdefault}{\mddefault}{\updefault}{\color[rgb]{0,0,0}$\alpha_{n-1}$}%
}}}}
\put(2619,-1368){\makebox(0,0)[lb]{\smash{{\SetFigFont{7}{8.4}{\rmdefault}{\mddefault}{\updefault}{\color[rgb]{0,0,0}$\alpha_{n}$}%
}}}}
\put(603,-1085){\makebox(0,0)[lb]{\smash{{\SetFigFont{7}{8.4}{\rmdefault}{\mddefault}{\updefault}{\color[rgb]{0,0,0}$\alpha_{2}$}%
}}}}
\put(237,-1368){\makebox(0,0)[lb]{\smash{{\SetFigFont{7}{8.4}{\rmdefault}{\mddefault}{\updefault}{\color[rgb]{0,0,0}$\alpha_{1}$}%
}}}}
\put(2565,-4287){\makebox(0,0)[lb]{\smash{{\SetFigFont{7}{8.4}{\rmdefault}{\mddefault}{\updefault}{\color[rgb]{0,0,0}$\alpha_{n}$}%
}}}}
\put(237,-1652){\makebox(0,0)[lb]{\smash{{\SetFigFont{7}{8.4}{\rmdefault}{\mddefault}{\updefault}{\color[rgb]{0,0,0}Every root $\alpha_{i}$ is of non-compact type.}%
}}}}
\put(237,-2594){\makebox(0,0)[lb]{\smash{{\SetFigFont{7}{8.4}{\rmdefault}{\mddefault}{\updefault}{\color[rgb]{0,0,0}Only the root $\alpha_{n}$ is of non-compact type.}%
}}}}
\put(237,-3537){\makebox(0,0)[lb]{\smash{{\SetFigFont{7}{8.4}{\rmdefault}{\mddefault}{\updefault}{\color[rgb]{0,0,0}Only the root $\alpha_{1}$ is of non-compact type.}%
}}}}
\put(237,-4624){\makebox(0,0)[lb]{\smash{{\SetFigFont{7}{8.4}{\rmdefault}{\mddefault}{\updefault}{\color[rgb]{0,0,0}Only the roots $\alpha_{1}$, $\alpha_{2}$ and $\alpha_{n}$ are of non-compact type.}%
}}}}
\end{picture}%

%% file: classification.bbl
\begin{thebibliography}{10}

\bibitem{Bal} V.\,K.~Balachandran, {\it Simple $L^{*}$-algebras of classical
      type}, Math. Ann. {180}, ({1969}), 205-219.

\bibitem{BRT} D.~Belti\c t\u a, T.~Ratiu, A.B.~Tumpach,  {\it
The restricted Grassmannian, Banach Lie-Poisson spaces, and
coadjoint orbits}, to appear in \textit{Journal of Functional
Analysis}.

\bibitem{Bes} A.\,L.~Besse, {\it Einstein Manifolds}, Ergebnisse
der Mathematik und ihre Grenzgebiete, Springer-Verlag, ({1986}).



\bibitem{Bou} N.~Bourbaki, {\it Groupes et Alg{\`e}bres de Lie},
{\'E}l{\'e}ments de Math{\'e}matiques, chap {4-5-6}, Masson
({1981}).

\bibitem{Lie} S\'eminaire Sophus Lie, {\it Th\'eorie des alg\`ebres de
  Lie}, \'Ecole normale sup\'erieure, Paris, {1955}.

\bibitem{Har2} P.~de la Harpe,  {\it Classification des
    $L^{*}$-alg{\`e}bres semi-simples r{\'e}elles s{\'e}parables},
    C.R. Acad. Sci. Paris, Ser. A {272} ({1971}), 1559-1561.

\bibitem{Hel} S.~Helgason,  {\it Differential Geometry and
  Symmetric Spaces}, Academic Press, New York, ({1962}).

\bibitem{Kau1} W.~Kaup,  {\it Algebraic
characterization of symmetric complex
    Banach manifolds}, Math. Ann. {228}, ({1977}), 39-64.

\bibitem{Kau2} W.~Kaup,  {\it {\"U}ber die Klassifikation der
  symmetrischen hermiteschen Mannigfaltigkeiten unendlicher Dimension
  I, II}, Math. Ann. {257}, ({1981}), 463-486,
  {262},  ({1983}), 57-75.


\bibitem{Nee2} K.-H.~Neeb,  {\it Highest weight representations and
    infinite-dimensional K{\"a}hler manifolds}, Recent advanceds in Lie
    theory (Vigo, 2000), 367-392, Res. Exp. Math., {25}, Heldermann,
    Lemgo, {2002}.

\bibitem{Ne02c}
 K.-H.~Neeb, {\it A Cartan-Hadamard theorem for Banach-Finsler
manifolds},  Geom. Dedicata {95} ({2002}), 115--156.


\bibitem{PS} A.~Pressley,  G.~Segal,  {\it Loop Groups},
Oxford Mathematical Monographs. Oxford (UK): Clarendon Press.
viii, 318 p. ({1988})

\bibitem{Schu1} J.\,R.~Schue,  {\it Hilbert space methods in the theory
    of Lie algebras}, Trans. Amer. Math. Soc. {95} ({1960}),  69-80.

\bibitem{Schu2} J.\,R.~Schue,  {\it Cartan decompositions for
    $L^{*}$-algebras}, Trans. Amer. Math. Soc {98}, ({1961}),  334-349.

\bibitem{Wol1} J.\,A.~Wolf,
{\it Fine structure of Hermitian Symmetric Spaces}, Symmetric
Spaces, short Courses presented at Washington Univ., pure appl.
Math. {8},
 ({1972}), 271-357.

\bibitem{Wol2} J.\,A.~Wolf,  {\it On the classification of
  Hermitian Symmetric Spaces}, J. Math. Mech., { 13}, ({1964}), 489-496.

\bibitem{Uns2} I.~Unsain,  {\it Classification of the simple real
    separable $L^{*}$-algebras}, J. Diff. Geom. {7}, ({1972}), 423-451.

\bibitem{Tum} A.\,B.~Tumpach,  {\it Vari\'et\'es k\"ahl\'eriennes et
hyperk\"ahl\'eriennes de dimension infinie}, Ph.D Thesis, \'Ecole
Polytechnique, Palaiseau, France, (july {2005}), electronic
version avaible on http://bernoulli.epfl.ch/hosted/tumpach.


\bibitem{Tum1} A.\,B.~Tumpach,  {\it Hyperk\"ahler structures and infinite-dimensional Grassmannians},
J. Funct. Anal. 243 (2007), 158-206.

\bibitem{Tum2} A.\,B.~Tumpach,  {\it Mostow Decomposition Theorem
for a $L^*$-group and Applications to affine coadjoint orbits and
stable manifolds}, preprint arXiv:math-ph/0605039, (May {2006}).

\bibitem{Tum3} A.\,B.~Tumpach,  {\it Infinite-dimensional hyperk\"ahler manifolds associated with
Hermitian-symmetric affine coadjoint orbits}, preprint
arXiv:math-ph/0605032, ({2006}).


\bibitem{Wur1} T.~Wurzbacher,  {\it Fermionic Second Quantization and the
    Geometry of the Restricted Grassmannian}, in Infinite-Dimensional
    K{\"a}hler Manifolds, DMV Seminar, Band {31}, Birkh{\"a}user, ({2001}).



\end{thebibliography}
